\documentclass{rspublic} 
\usepackage{graphicx} 
 
\begin{document} 
 
\title
[Swift observations of GRBs]
{Using 
Swift observations of prompt and afterglow emission to classify GRBs}

\author[P.T. O'Brien, R. Willingale]{Paul T. O'Brien, Richard Willingale} 
 
\affiliation{Department of Physics and Astronomy, University of Leicester,  
University Road, Leicester, LE1 7RH, UK}

\maketitle

\begin{abstract}{Gamma-ray bursts, black holes, X-ray} 
 
We present an analysis of early BAT and XRT data for 107 gamma--ray
bursts (GRBs) observed by the {\it Swift} satellite. We use these data
to examine the behaviour of the X-ray light curve and propose a
classification scheme for GRBs based on this behaviour. As found for
previous smaller samples, the earliest X-ray light curve can be well
described by an exponential which relaxes into a power law, often with
flares superimposed. The later emission is well fit using a similar
functional form and we find that these two functions provide a good
description of the entire X-ray light curve. For the prompt emission,
the transition time between the exponential and the power law gives a
well-defined timescale, T$_p$, for the burst duration. We use T$_p$,
the spectral index of the prompt emission, $\beta_p$, and the prompt
power law decay index, $\alpha_p$ to define four classes of burst:
short, slow, fast and soft. Bursts with slowly declining emission have
spectral and temporal properties similar to the short bursts despite
having longer durations. Some of these GRBs may therefore arise from
similar progenitors including several types of binary system. Short
bursts tend to decline more gradually than longer duration bursts and
hence emit a significant fraction of their total energy at times greater
than T$_p$. This may be due to differences in the environment or the
progenitor for long, fast bursts.
 
\end{abstract}

\section{Introduction}

Gamma-ray bursts (GRBs) were originally discovered as short flashes of
gamma-rays seen from a random, non-repeating location on the sky
(Klebesadel {\it et al.} 1973). GRBs are now known to be
extragalactic, many with large redshifts (Costa {\it et al.} 
1997). During the prompt phase a GRB is intrinsically the most
luminous single object in the Universe.  Understanding the observed
emission in the first few minutes to days following a burst is crucial
in determining the nature of the energy source and the
progenitor. Since its launch in November 2004, the {\it Swift}
satellite (Gehrels {\it et al.} 2005) has provided a unique set of
early-time X-ray light curves. Combining data from the Burst Alert
Telescope (BAT) and the X-ray Telescope (XRT) allows for a
determination of the temporal and spectral properties of a burst from
the initial trigger out to days or even weeks.
 
It has become conventional to place GRBs into classes dependent on the
observed duration of the prompt emission seen in the gamma-ray
band. The T$_{90}$ parameter is the timescale over which 90\% of the
gamma-rays were detected. Those GRBs for which T$_{90}$ exceeds 2
seconds are described as ``long'' whereas those shorter than 2 seconds
are described as ``short'' (e.g. Kouveliotou {\it et al.} 1993). The
short bursts tend to have harder gamma-ray spectra than the long
bursts. The observed flux can be represented as a
function of time and frequency using $f_\nu\propto \nu^{-\beta}
t^{-\alpha}$, where $\beta$ is the spectral index and $\alpha$ is the
temporal index. The spectral index is related to the photon index
$\Gamma$ by $\Gamma = \beta + 1$.
 
In this paper we summarise the X-ray observational results from {\it
Swift}, following on from previous work (e.g. Tagliaferri {\it et al.} 
2005; Nousek {\it et al.} 2006; O'Brien {\it et al.} 2006a).  These
papers and many others have shown that GRBs display a wide variety of
phenomena during the few hours to days after discovery but that there
are some common patterns emerging. It is already clear that a number
of different emission components probably contribute to the observed
flux, some due to processes in the luminous central engine or
relativistic jet and some due to interaction of the jet with the
surrounding environment. It is not our intention here to review all
the possible emission processes but rather provide an overall view of
the observed temporal and spectral behaviour of the largest {\it
Swift} GRB sample yet studied and use them to suggest a GRB
classification scheme.

We combine data from the BAT and XRT on {\it Swift} using a technique
described in O'Brien {\it et al.} (2006a) but extended to provide a
better description of the latter-time emission. We include data for
107 GRBs discovered by {\it Swift} up to 2006 August 1 for which
prompt XRT observations were obtained. Here we discuss the behaviour
concentrating on the first few hours to a day or so. In a companion
paper (Willingale \& O'Brien 2006) we discuss the latter emission and
examine the correlations of observed properties with luminosity. This
work is discussed in detail in Willingale {\it et al.} (2006).  The
sample is dominated by long bursts but includes a number of short
bursts and probes the full range in observed fluence and
redshift for GRBs detected by {\it Swift}.
 
\begin{figure}
\begin{center}
  \includegraphics[angle=0,width=.5\textheight]{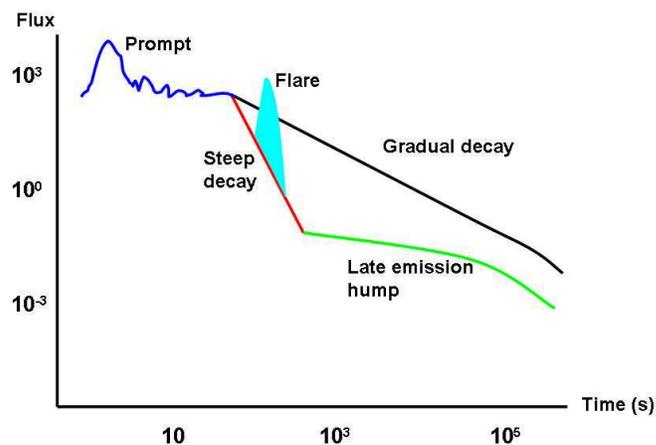}
\end{center}
  \caption{A schematic view of the early GRB X-ray light
  curve. Following the prompt emission, which typically lasts a few
  10s of seconds, the decay tends to follow one of two paths: (i) a
  steep decay, during which the flux can fall by several orders of
  magnitude, followed by a shallower, ``late emission hump'' starting
  at $\sim 10^3$s; or (ii) a more gradual decay. Either decay path can
  end with a break at $>10^4$s to a steeper decay. X-ray flares can
  occur during either decay path, most prominently during the first
  hour.}
\label{figure1} 
\end{figure}

\section{BAT and XRT analysis} 
 
The BAT and XRT data were processed using the standard {\it Swift}
analysis software. For the BAT, light curves and spectra were extracted
over the 15--150 keV band. Spectral indices ($f_\nu\propto
\nu^{-\beta}$), were derived by fitting power laws over the T$_{90}$
period. Power law fits were also used to parameterise the XRT spectra
over the 0.3--10 keV band.  Most of the GRBs (59 out of 107) show
evidence for excess absorption above the Galactic column.
 
The spectra extracted from the XRT data are usually best fitted by a
softer power law than the BAT data (O'Brien {\it et al.} 
2006a). Hence, to construct unabsorbed, 0.3--10 keV flux light curves
we converted the BAT count rates into unabsorbed fluxes by
extrapolating the BAT data to the XRT band using a power law model
with an absorbing column derived from the XRT data and a spectral
index which is the mean of the XRT and best-fit BAT spectral indices.
The XRT count rates were converted into unabsorbed fluxes using the
early XRT power law model.
 
\section{The observed early temporal and spectral shape} 
 
The variety in observed GRB light curves is shown schematically in
figure 1 (O'Brien {\it et al.} 2006b). and can be summarized as:

\begin{enumerate}

\item There is a ``prompt'' phase which includes the emission seen emitted
during the burst. With {\it Swift} this emission is initially seen by
the BAT but can also be detected by the XRT if the burst is long
enough to last until completion of the first slew on to the
target. Most bursts observed by {\it Swift} typically have a spectral
index of 0--2 in the 15--150 keV band during the prompt phase.

\item The burst is followed by a power law decline. The observed
temporal index, $\alpha$, during this phase can be very large (up to
$\approx 6$) but clusters around 2 (section 5). In a significant
minority of GRBs the initial decay is more gradual with $0.5 \le
\alpha \le 1.5$.   The spectral index in the 0.3--10
keV band during the initial power law decay is usually in the range
0.5--2.5. In the fitting procedure, described below, we fit both the
burst and the initial decline as one ``prompt phase'' denoted by
subscript $p$ in fitted parameters. Emission following that phase is
denoted by subscript $a$ (for afterglow) in fitted parameters.  The
wide range in the prompt temporal and spectral indices imply that
several emission processes may be involved. O'Brien {\it et al.} 
(2006a,b) suggested that a combination of ``high latitude'' emission
(Kumar \& Panaitescu 2000) and afterglow emission (e.g. M\'esz\'aros
\& Rees 1997) can explain the early emission. This conclusion also applies
for the larger GRB sample used here.

\item For those bursts which initially decline steeply, the decay breaks
to a shallower rate, typically within the first hour, such that $0.5\le
\alpha \le 1.5$. This ``late emission hump'', modelled as an additional 
component (section 4), can last for up to $\sim 10^{5}$s before
breaking again to a steeper decay. Although faint in flux, this phase
lasts so long that it has a significant fluence, so it is
energetically significant. In fact the late emission has at most a
fluence equal to that of the prompt phase (O'Brien {\it et al.} 
2006a). 

\item Limited statistics make quantifying later phases difficult, but both
the initially steeply declining bursts and those that decline more gradually
can show a late temporal break (typically at $10^4$ -- $10^5$s) to a
steeper decay. These late breaks are not seen in all GRBs --- some decay
continuously beyond $10^6$s until they fade below the {\it Swift} XRT
detection limit. The behaviour of this phase is described in Willingale
\& O'Brien (2006) and Willingale {\it et al.} (2006).

\item X-ray flares are seen in the first few hours for around half of the
GRBs observed by {\it Swift} , and occur in GRBs which decline rapidly
or gradually. The majority of these flares are only detected in the
XRT but in some bright, long bursts flares are observed simultaneously
with the BAT.  Strong spectral evolution can be observed in some
cases. Most of the X-ray flares are energetically small, but a few are
very powerful with a fluence comparable to that of the prompt phase
(e.g. Burrows {\it et al.} 2005). Late flares are also occasionally
seen. Flares are excluded in the fitting procedure used here and we do not
consider them further in this paper.

\item The X-ray light curves for short bursts have been less well studied
by {\it Swift} as they are fainter (on average) and {\it Swift} has detected
fewer examples. The {\it Swift} short bursts are included in our GRB sample.

\end{enumerate}

\begin{figure}
\begin{center}
  \includegraphics[angle=-90,width=.6\textheight]{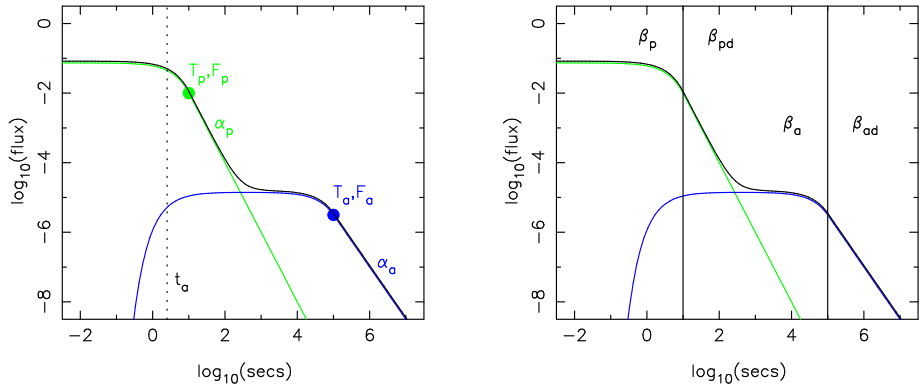}
\end{center}
  \caption{The left-hand panel shows the functional form of the fit to
  the X-ray light curve. Two components, denoted by subscripts $p$ and
  $a$ for prompt and afterglow respectively, are shown. For the
  spectral fits (right-hand panel) subscript $d$ is also used to
  distinguish between the spectral shape during the exponential and
  the power law decay. The exponential relaxes into a power law at
  time $T$ where the flux is $F$. The parameter $\alpha$ controls both
  the time constant of the exponential ($T/\alpha$) and the temporal
  decay index of the power law. For the second, afterglow component,
  which describes the late emission hump, an initial rise has been
  introduced which for fitting purposes occurs at the transition time
  of the prompt component. Changing the shape of this rise to a
  constant value at early times has no significant effect on the
  fits.}
\label{figure2} 
\end{figure}

The behavioural pattern of prompt emission followed by a steep X-ray
decay and then a shallow decay has been characterised as the
``canonical GRB light curve'' (Nousek {\it et al.} 2006).  This
pattern is seen in the majority of GRBs (perhaps three-quarters or
so), but not in all. To try and understand the different phases we have
developed a parameterised functional fitting procedure summarised below.

\begin{figure}
\begin{center}
  \includegraphics[angle=0,width=.5\textheight]{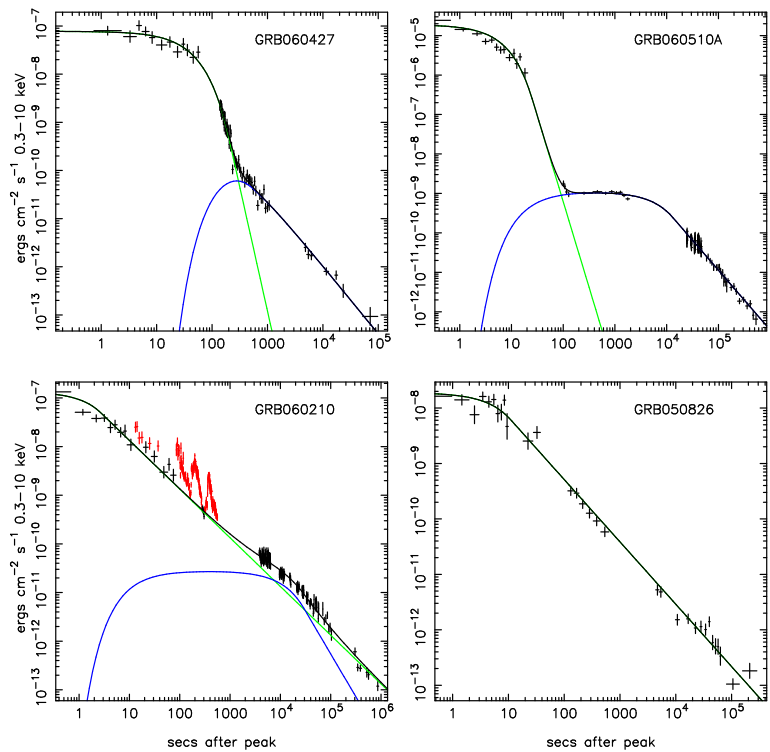}
\end{center}
  \caption{Example fits to the X-ray light curves.}
\label{figure3} 
\end{figure}

\section{The parameterised temporal and spectral behaviour}

In order to compare the different GRB light curves we have followed
and expanded the procedure outlined in O'Brien {\it et al. } 
(2006a). There we found that the initial prompt emission was described
well by an exponential which relaxes into a power law. Here we extend
that concept to fit the prompt emission and the late emission hump
using a similar function.
 
The flux light curve ($f(t)$) for both the prompt and later emission
are well fitted by the same functional form:
\begin{equation}
f_{p}(t)=F_{p}
\exp\left(\alpha_{p}-\frac{t\alpha_{p}}{T_{p}}\right)
\exp\left(\frac{-t_{p}}{t}\right)
,\:\:t<T_{p}
\label{eq2}
\end{equation}
\begin{equation}
f_{p}(t)=F_{p} \left( \frac{t}{T_{p}}\right)^{-\alpha_{p}}
\exp\left(\frac{-t_{p}}{t}\right)
,\:\:t\ge T_{p}
\label{eq3}
\end{equation}

The subscript $p$ refers to the prompt component (figure 2). The
transition from the exponential to the power law occurs at the point
$(T_{p},F_{p})$ where the two functional sections have the same value
and gradient. A similar pair of equations are used to fit the late
emission hump with an ``afterglow'' component, denoted by subscript
$a$. Thus, we have fitted the X-ray light curves of all of the GRBs
using two components of the form $f(t)=f_{p}(t)+f_{a}(t)$. Details of
the fitting procedure are given in Willingale {\it et al.} (2006). The
fitted components are shown schematically in figure 2 along with the
corresponding spectral indices. 

Four example fits are shown in figure 3. The top panels show examples
of the most common type of burst in which there is a detectable second
component. The initial decay is very steep for GRB060427 ($\alpha_p
\sim 5$) although it only lasts for a very short period of time. Similarly
for GRB060510A.  In the bottom-left panel the afterglow component is
less prominent in GRB060210, but forms a bump in the light curve. In
this burst the initial decay is quite shallow ($\alpha_p \sim 1$). The
numerous small flares in GRB060210 do not appear to disturb the
overall decay. The afterglow component is not detectable in GRB050826
(bottom-right panel).

Using the fitting procedure, we obtain results for the larger GRB
sample which are consistent with those presented in O'Brien {\it et
al.}  (2006a,b). Rather than restate the full results of that analysis,
for the rest of this paper we concentrate on the possibility of using the
fitted parameters to help classify GRBs.

\section{Classifying a GRB}

Traditionally GRBs have been classified into short and long bursts
based primarily on T$_{90}$. Those below 2 seconds are classified as
short and it has been noticed that they tend to have harder prompt
spectra (Kouveliotou {\it et al.} 1993). In the {\it Swift} era we
should examine if additional parameters, such as how fast the burst
decays, may provide additional information. It has already been
noticed that some bursts which appear long in terms of T$_{90}$ may be
more related to short bursts in terms of their host galaxies
(e.g. Barthelmy {\it et al.} 2005) or in terms of the observed
inter-band time-delays (e.g. Gehrels {\it et al.} 2006). Unfortunately
data such as the nature of the GRB host are not always available.
Here we examine if use of the T$_p$ parameter, the prompt spectral
index, $\beta_p$, and the early X-ray decay rate, $\alpha_p$, can be
used to provide a classification scheme.

\begin{figure}
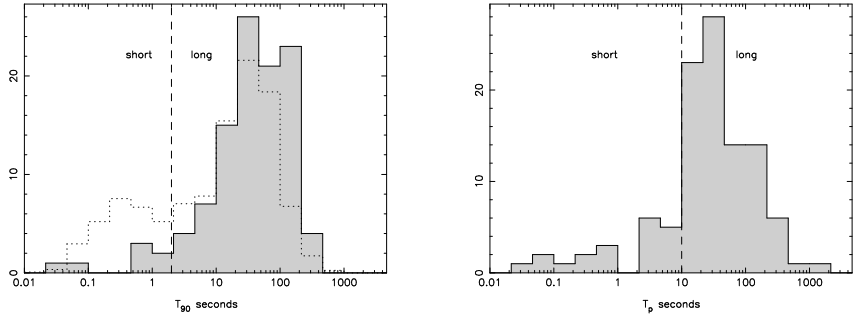

\begin{center}
  \includegraphics[angle=-90,width=.3\textheight]{figure4a.eps}
\includegraphics[angle=-90,width=.3\textheight]{figure4b.eps}
\end{center}
  \caption{The {\it Swift} T$_{90}$ and T$_p$ distribution. The dotted
  distribution in the left-hand panel is the T$_{90}$ distribution for
  {\it BATSE} normalised to the number of {\it Swift} bursts. In
  either panel a possible division in GRB type is plotted as a vertical dashed
line corresponding to 2 seconds for T$_{90}$ and 10 seconds for T$_p$. 
}
\label{figure4} 
\end{figure}
 
The transition time between the prompt exponential and power law decay can 
be used to define a measure of burst length, T$_p$. This uses information 
from both the BAT and XRT instruments and rather than a measure of how 
long the gamma-ray instrument detected the burst instead is a measure 
based on the shape of the light curve.  For many burst T$_{p}$ is 
comparable to T$_{90}$ (O'Brien {\it et al.} 2006a,b) but it can be 
considerably shorter or longer.

In figure 4, the left-hand panel compares the T$_{90}$ distribution
for the {\it BATSE} (Paciesas {\it et al.} 1999) and {\it Swift}
samples including the conventional division at 2 seconds between short
and long bursts. Compared to {\it BATSE}, the {\it Swift} sample
contains significantly fewer short bursts as a fraction of the total,
possibly due to the different spectral range and trigger system
(e.g. Band 2006).  It is not clear there is a division between the two
populations in the {\it Swift} sample. The possibility of a more
continuous distribution is further emphasised if we use the T$_p$
distribution, as shown in the right-hand panel of figure 4. In that
panel we define a division at 10 seconds which we propose as it is close
to the peak in the T$_p$ distribution and defines a tail of shorter values.

\begin{figure}
\begin{center}
  \includegraphics[angle=-90,width=.6\textheight]{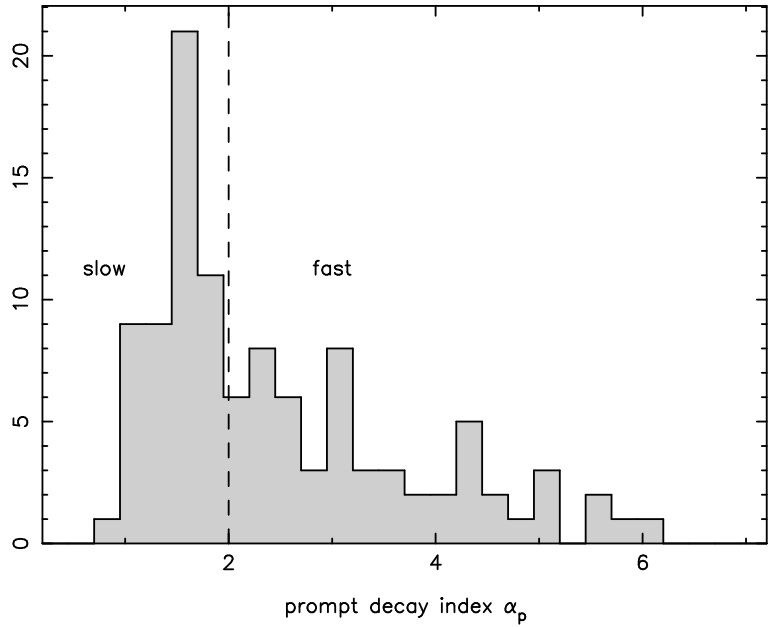}
\end{center}
  \caption{The distribution of initial decay index, $\alpha_p$, during
  the prompt phase. The sample has been divided into two groups --
  slow and fast -- either side of $\alpha_p = 2$.}
\label{figure5} 
\end{figure}

There is a wide range in the observed initial decay rate, $\alpha_p$,
following the burst (figure 5). There are some GRBs extending up to
very high rates of decay, although such rates are maintained for only
a short period of time. The decay index distribution shows a peak
close to $\alpha_p = 2$ and a tail to higher values, so we have divided
the sample into ``slow'' and ``fast'' decays at $\alpha_p =2$.  This
divides the population approximately in half.

\begin{figure}
\begin{center}
  \includegraphics[angle=-90,width=.4\textheight]{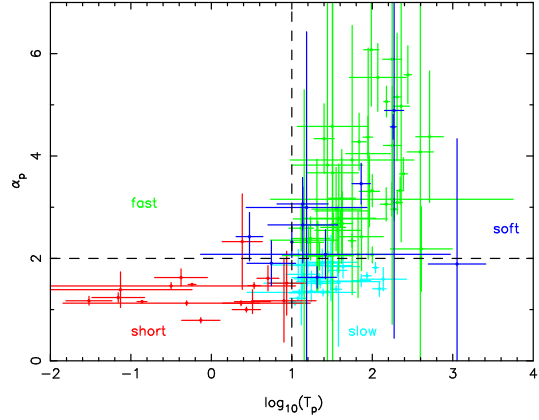}
\end{center}
  \caption{The correlation of the prompt decay rate, $\alpha_p$, with the 
burst duration T$_p$. The bursts are divided at $\alpha_p = 2$ and T$_p 
=10$ seconds as described in the text.}
\label{figure6} 
\end{figure}

\begin{figure}
\begin{center}
  \includegraphics[angle=-90,width=.5\textheight]{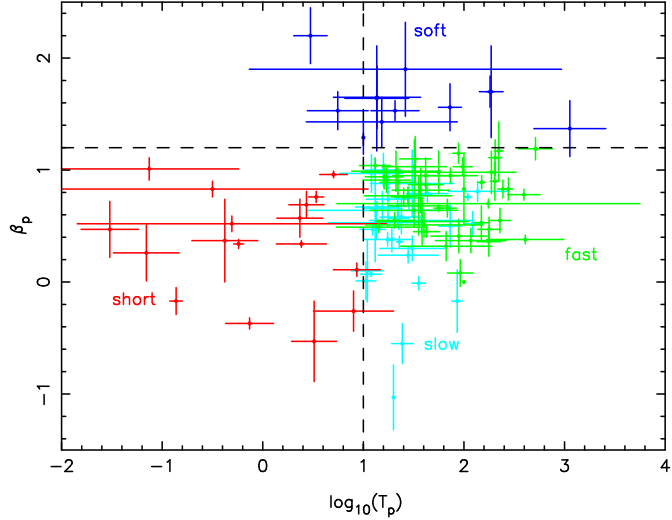}
\end{center}
  \caption{The correlation of the prompt spectral index, $\beta_p$, with 
the 
burst duration T$_p$. The bursts are divided at $\beta_p = 1.2$ and T$_p 
=10$ seconds as described in the text.}
\label{figure7} 
\end{figure}

Examining the correlation of $\beta_p$ and $\alpha_p$ with T$_p$
(figures 6 and 7) shows that the shorter duration bursts tend to be
harder, as found previously, but also decay more gradually following
the burst. From these figures it can be seen that a value of $\beta_p
= 1.2$ combined with the previously discussed T$_p = 10$ and $\alpha_p
= 2$ sub-divides the GRBs into well-defined sectors of figures 6 and
7.  We therefore propose a sub-division of GRBs into four groups based
around cuts at T$_p = 10$ seconds, $\beta_p = 1.2$ and $\alpha_p =
2$. Using both spectral and temporal properties to classify GRBs is
reasonable as these properties should depend in part on the nature of
the progenitor and its surrounding environment. Using the observed
properties may lead to a closer understanding of the various types of
progenitor. The four GRB groups are defined as:

\begin{enumerate}

\item Short bursts with T$_p < 10$ seconds and $\beta_p 
< 1.2$.

\item Slow bursts with T$_p > 10$ seconds, $\alpha_p < 2$ and $\beta_p 
< 1.2$.

\item Fast bursts with T$_p > 10$ seconds, $\alpha_p > 2$ and $\beta_p 
< 1.2$.

\item Soft bursts with T$_p$ and $\alpha_p$ of any value and 
$\beta_p 
> 1.2$.

\end{enumerate}

Although, as in all such schemes, such as using T$_{90}$, one can
argue about precisely what exact division in parameter space to use,
this classification method does separate the two main groups of short
and fast (long) bursts, but at a larger value of duration (T$_p =10$
seconds) than traditionally used for T$_{90}$. It also defines two
further groups, one of which --- the slow bursts -- have spectral and
temporal decay properties which overlap with the short bursts. The
soft bursts are define solely in terms of spectral shape, are similar
to X-ray flashes and overlap mostly with the longer duration bursts.

\begin{figure}
\begin{center}
  \includegraphics[angle=-90,width=.5\textheight]{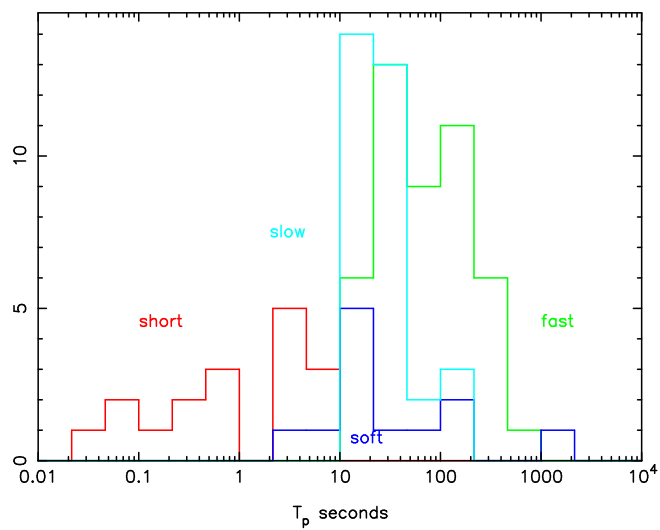}
\end{center}
  \caption{The distribution of the four GRB groups with T$_p$.}
\label{figure8} 
\end{figure}

The distribution of the four groups of GRB as a function of duration
T$_p$ is shown in figure 8. It is interesting that the slow bursts
form a clear peak. Of course this peak is truncated at the short
duration end by the definition of short bursts applied here, but falls
rapidly at long durations in contrast to the fast decay bursts. It is
thus tempting to suggest that a fair fraction of the slow bursts may
be longer-duration variants of the short bursts and hence may arise
from similar progenitors. The fast, longer-duration GRBs may arise
from a different progenitor if the rate of decay of the
X-ray emission is related to the local environment. The soft GRBs are
located towards the centre of the distribution. The long-duration soft
GRB at the far-right of the distribution is GRB060218, an unusual GRB
in many respects (Campana {\it et al.} 2006).

The classification scheme outlined here is an extension of that
previously used but can be improved by better statistics and by
including other parameters where available, including the inter-band
time-delay (Norris 2002). Currently, some bursts which appear long in
terms of T$_{90}$ nevertheless have other properties which appear to
place them in the subset of GRBs which may not be due to collapsars
(e.g. Barthelmy {\it et al.} 2005; Gehrels {\it et al.} 2006). It has
been usual to suppose that binaries may not produce long-duration
bursts, and indeed this seems unlikely for neutron-star
binaries. However, other binaries, such as neutron-star black hole
systems or white-dwarf neutron-star systems (King 2006), may be
capable of producing emission over longer intervals if one of the
binary pair is fragmented and accreted over a relatively long
period. Such systems could help explain the fact that the ``short'' bursts
which decay slowly actually emit a large fraction ($\ge 50$\%) of
their energy at times greater than T$_p$. If a scheme involving all of
the observed prompt parameters can be developed it may be possible to
disentangle the various progenitor types and truly classify GRBs.

\section{Acknowledgments} The authors gratefully acknowledge funding for 
{\it Swift} at the University of Leicester by PPARC, in the USA by NASA 
and in Italy by ASI.

\end{document}